\documentclass[twocolumn,prl]{revtex4}
\usepackage{graphicx}
\usepackage{dcolumn}
\usepackage{bm}
\usepackage{amsmath}
\usepackage{amsfonts}
\usepackage{psfrag}
\usepackage{color}

\newcommand{\Eq}[1]{Eq. (\ref{#1})}
\newcommand{\ca}[2]{\tilde{c}_{#1,\mathbf{#2}}}
\newcommand{\cc}[2]{\tilde{c}^{\dagger}_{#1,\mathbf{#2}}}
\newcommand{\cam}[2]{c_{#1,\mathbf{#2}}}
\newcommand{\ccm}[2]{c^{\dagger}_{#1,\mathbf{#2}}}
\newcommand{\ma}[1]{\tilde{a}_{\mathbf{#1}}}
\newcommand{\mc}[1]{\tilde{a}^{\dagger}_{\mathbf{#1}}}
\newcommand{\mam}[1]{a_{\mathbf{#1}}}
\newcommand{\mcm}[1]{a^{\dagger}_{\mathbf{#1}}}
\newcommand{\fa}[1]{\delta_{\mathbf{#1}}}
\newcommand{\fc}[1]{\delta^{\dagger}_{\mathbf{#1}}}

\begin{document}
\title{Quantum phases of a multimode bosonic field coupled to flat electronic bands
}
\author{Simone \surname{De Liberato}$^{1,2}$}
\author{Cristiano Ciuti$^2$}
\affiliation{$^1$School of Physics and Astronomy, University of Southampton, Southampton, SO17 1BJ, United Kingdom}
\affiliation{$^2$Laboratoire Mat\'eriaux et Ph\'enom\`enes Quantiques, Universit\'e Paris  Diderot-Paris 7 and CNRS, UMR 7162, 75013 Paris, France} 
 
\begin{abstract}
We investigate the quantum phases of systems in which a multimode bosonic field is coupled to the transitions between two flat electronic bands. In the literature, such systems are usually modeled using a single or multimode Dicke model, leading to the prediction of superradiant quantum phase transitions for large enough couplings.
We show that the physics of  these systems is remarkably richer than previously expected, with the system continuously interpolating between a Dicke model exhibiting a superradiant quantum phase transition and a quantum Rabi model with no phase transition.
\end{abstract}

\maketitle

In cavity quantum electrodynamics  experiments the photon confinement increases the light-matter coupling, eventually allowing for a single photon to be absorbed and re-emitted multiple times before escaping out of the cavity. It is the hallmark of the strong coupling regime, realized for the first time with Rydberg atoms in superconducting cavities \cite{Haroche}. Still, while the strong coupling regime is observable in these systems thanks to the impressive quality factors of superconducting cavities, the strength of the dipolar coupling between a single Rydberg atom and the electromagnetic field, quantified by its quantum Rabi frequency, $\Omega$, is bounded by fundamental reasons to be a small fraction of the frequency of the transition, $\omega_{12}$ \cite{Devoret07}.

This bound can be lifted if photons couple to collective superradiant electronic excitations \cite{Dicke54}: $N$ identical, coherently excited dipoles, behave as a single super dipole  $\sqrt{N}$ times larger. Such a phenomenon opens to the possibility  to experimentally achieve the ultrastrong coupling regime, in which the vacuum Rabi frequency becomes of the same order of the bare frequency of the electronic excitations, and the normalized coupling, $\frac{\Omega}{\omega_{12}}$, is thus of order one \cite{Ciuti05,Anappara09,Todorov10,Niemczyk10,Hagenmuller10,Scalari12}. 
As in this regime the interaction energy  is one of the dominant energy scales of the problem, it becomes possible to observe fascinating new physical phenomena, ranging from  quantum vacuum radiation  \cite{DeLiberato07,Auer12,Dodonov12} to superradiant quantum phase transitions (QPT)\cite{Emary03,Lambert04,Nataf10,Baumann10,Strack11,Bastidas12,Bhaseen12}.

These systems are usually studied using the Dicke model  \cite{Dicke54}, describing $N$ two level atoms identically coupled to the electromagnetic field. This model is based on the assumption that the interaction with the photonic field can only induce transitions between levels in the same atom. No transitions between different atoms are allowed.
While this assumption is well justified for dilute atomic clouds, this may not be the case in solid state systems \cite{Todorov10,Scalari12} or in synthetic many-body systems \cite{Baumann10,Ruostekoski09,Jo12,Wu07}, where the two level systems can form, instead, two electronic bands. In this case, each state of the lower band can be coupled, through different photonic modes, to different final states in the upper band. 

The Dicke model can only give an approximate description of these band models, as it can be grasped by a simple dimensional argument. The Hilbert space of $N$ electrons in a Dicke model has dimension $d_{\text{DM}}=2^N$ ($N$ two level systems).
For the same $N$, the electronic Hilbert space in a  band model, in which electrons can  jump from any state of the lower band to any state of the upper one, is instead $d_{\text{FBM}}=\left (\begin{array}{c} 2N \\ N \end{array} \right )$ ($N$ electrons spread over $2N$ possible states),  that tends to $\frac{4^N}{\sqrt{\pi N}}$ for large $N$. The physical Hilbert space  is therefore much larger than the one of the Dicke model and, as we will see, this translates in a much richer physics.

%Usual, low intensity spectroscopic experiments, elicit a linear response, in which each mode composing the probe beam can be treated independently. As a consequence, the response of the system is well described by the Dicke model, in which only a single transition is considered \cite{Hagenmuller10,Anappara09,Todorov10,Scalari12}. 
%One remarkable exception to the aforementioned argument is the case in which the light-matter coupling strength is larger than the critical value for the system to undergo a superradiant QPT. In this case, the ground state population of the excited electronic band becomes macroscopic, the response of the system nonlinear, and we expect to observe physics beyond that of the Dicke model.

In this Letter, in order to investigate this new and complex physics, we will study a  model describing a multimode bosonic field coupled to two flat electronic bands, separated by a bandgap $\omega_{12}$. 
This particular choice is motivated by the fact that flat band models, that can be engineered using a number of different systems, from two dimensional electron gases \cite{Hagenmuller10,Scalari12} and graphene under magnetic field \cite{Novoselov07,Hagenmuller12}, to atoms in optical lattices \cite{Wu07,Ruostekoski09,Jo12} and topological insulators \cite{Tang11,Neupert11,Sun11,Wang11}, are characterized by density of states peaked around the badgap, leading to the possibility to observe extremely large light-matter couplings \cite{Hagenmuller10,Scalari12,Hagenmuller12}.

We will prove that, modifying the spectrum of the bosonic field, it is possible to continuously tune the physics from the usually expected Dicke model to an ultrastrongly coupled quantum Rabi model \cite{Rabi37,Braak11,Solano11,Ballester12}. Hence, the engineering of the bosonic spectrum can allow to simulate different classes of superradiant models, with a dramatic impact on the phase diagram.

\begin{figure}[h!]
\begin{center}
\includegraphics[width=7.5cm]{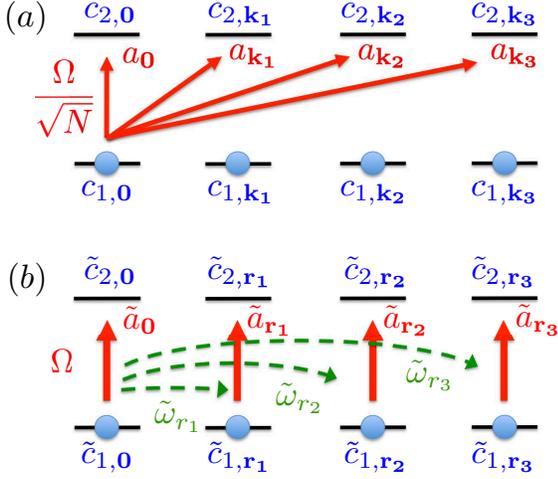}
\caption{\label{sketch}
(a) Sketch of the considered model with flat bands in momentum space: $N$ electrons can occupy two parallel bands, with $N$ states each, and each state in the first band couples to every state of the second one through a different bosonic mode, with a coupling constant $\frac{\Omega}{\sqrt{N}}$.
(b) Real space picture:  each state in the first band couples with only one state in the second band, through a
local bosonic field, with a coupling strenght $\Omega$. The different local bosonic modes at sites $\mathbf{r}$ and $\mathbf{r'}$ are further coupled between them by the coupling constant $\tilde{\omega}_{\lvert \mathbf{r-r'} \lvert}$.
}
\end{center}
\end{figure}

We consider two flat bands of states with $N$ modes each, with the Fermi level in the bandgap.  The states in each band are indexed by a general $n$-dimensional momentum, $\mathbf{k}$, with 
$k\in \lbrack -k_0, k_0\rbrack$,  and periodic boundary conditions.
Each state in the lower band, $\cam{1}{k}$, is coupled to each state of the upper band, $\cam{2}{k'}$, by the appropriate momentum-conserving bosonic mode, $\mam{q}$, with $\mathbf{q}=\mathbf{k'-k}$, through a coupling constant, $\Omega$, independent from $\mathbf{q}$ (see Fig. \ref{sketch} (a) for a schematic representation of the model).
Such a model is of course an idealization, and the physics of a real implementation could be modified by the presence of  cut-offs and inhomogeneities, due both to the  Fermi surface and to the wavevector dependency of the coupling constant.  
Still, this idealized model allows us to show, in a clear and paradigmatic way, how the presence of extra electronic transitions, coupled to different modes of the bosonic field, can profoundly modify  the physics with respect to the well-known Dicke model.

The Hamiltonian of the system, in momentum space, thus reads
 \begin{eqnarray}
 \label{H}
H^{\text{FBM}}&=&\sum_{\mathbf{q}}\omega_{q} \mcm{q}\mam{q}+\sum_{\mathbf{k}} \omega_{12} \ccm{2}{k}\cam{2}{k}\\
&+&\frac{\Omega}{\sqrt{N}} \sum_{\mathbf{k,q}} (\ccm{2}{k+q}\cam{1}{k}+\ccm{1}{k}\cam{2}{k-q})(\mcm{-q}+\mam{q}),
\nonumber
\end{eqnarray}
where $\omega_q$ is the frequency of the mode $\mam{q}$, that we will take to be isotropic, and we have set, here and in the rest of this paper, $\hbar=1$. 
As we have anticipated in the introduction,  the behavior of the Hamiltonian in \Eq{H} is controlled by the dispersion of the bosonic field, $\omega_q$. 
It is thus practical to introduce three quantities describing its average, $\bar{\omega}=\frac{1}{N}\sum_{\mathbf{q}} \omega_q$,
its half width, $\Delta=\bar{\omega}-\omega_m$, and the energy of the lower lying  mode, $\omega_m=\min_{\mathbf{q}} \, \omega_q$.

Defining the Fourier transform $\tilde{h}_{\mathbf{r}}$, of a quantity $h_{\mathbf{q}}$, as $h_{\mathbf{q}}=\frac{1}{\sqrt{N}}\sum_{\mathbf{r}} \tilde{h}_{\mathbf{r}} e^{i{\mathbf{qr}}}$ (${\mathbf{r}}$ is thus a point on an $n$-cubic lattice of side $\frac{2\pi}{k_0}$), we can rewrite the Hamiltonian in \Eq{H} in real space as
\begin{eqnarray}
\label{HF}
H^{\text{FBM}}&=&\sum_{\mathbf{r}}\bar{\omega} \mc{r}\ma{r}+ \omega_{12} \cc{2}{r}\ca{2}{r}\\&&+\Omega (\cc{2}{r}\ca{1}{r}+\cc{1}{r}\ca{2}{r})(\mc{r}+\ma{r})+\sum_{\mathbf{r'\neq r}}\frac{\tilde{\omega}_{\lvert \mathbf{r-r'}\lvert}}{\sqrt{N}} \mc{r}\ma{r'}\nonumber,
\end{eqnarray}
where we have exploited the fact that $\bar{\omega}=\frac{\tilde{\omega}_0}{\sqrt{N}}$.

Restricting the Hilbert space to states in which only one electron is present at each site $\mathbf{r}$, we can map the matter part at each site to a two level system. This is equivalent to neglect pairs of sites, one empty and one doubly occupied, completely decoupled from the electromagnetic field. This simplification can be justified, {\it a posteriori}, by the fact that the energy of such a pair, $\omega_{12}$, is much higher than the energies we will find for two interacting, singly occupied sites.

Hamiltonian in \Eq{HF} is reminiscent of a  Rabi-Hubbard model \cite{Tureci12}, describing a lattice of two-level systems, each coupled to a local bosonic mode of frequency $\bar{\omega}$, with a superradiantly enhanced coupling strength $\Omega$. The main difference between our model and the Rubi-Hubbard model described in Ref. \cite{Tureci12} is that in \Eq{HF} the coupling between different sites is not limited to nearest neighbors, but all the sites are coupled between them by complex coupling constants.  The coupling of two sites at distance $\mathbf{r}$ being given by the $r$-component of  the Fourier transform of the boson dispersion $\tilde{\omega}_r$ (see Fig. \ref{sketch} (b) for a schematic representation of the model in real space). 

For a flat bosonic dispersion, $\Delta=0$, $\omega_q=\omega_m=\bar{\omega}$, and  the only nonvanishing Fourier component corresponds to $r=0$.
Different sites thus completely decouple and in this limit the Hamiltonian in \Eq{HF} exactly maps on the sum of  $N$ identical and decoupled quantum Rabi models
\begin{eqnarray}
\label{HR}
H^{\text{QRM}}&=&{\omega}_{m} \mc{}\ma{}+ \frac{\omega_{12}}{2} \sigma^z+\Omega \sigma^x(\mc{}+\ma{}),
\end{eqnarray}
where $\sigma^j$, $j\in \lbrack x,y,z \rbrack$ are Pauli matrices. 
Notice that in \Eq{HR}, and in all the rest of this Letter, we discard
the constant terms from the Hamiltonian and neglect the site index for fully local and site-independent Hamiltonians.

This exact mapping has two, far-reaching consequences.
On one side, even if \Eq{HR} describes the physics of a single two level system, the light-matter coupling is enhanced by the superradiant factor $\sqrt{N}$.
Using flat bosonic dispersions, like the ones that can be engineered with photonic crystals, a flat band model can thus be used to simulate an ultrastrongly coupled quantum Rabi model, opening
to the possibility to observe new physics, as quantum vacuum radiation \cite{DeLiberato09} or the Bloch-Siegert shift \cite{Forn10}.
On the other side, while the Dicke model presents a phase transiton, when the coupling $\Omega$ exceeds a certain critical value \cite{Dicke54,Emary03,Lambert04}, the quantum Rabi model does not.   We thus proved that the Hamiltonian in \Eq{H} is in not always in the same universality class of the Dicke model. 

In the non-flat dispersion case however, the phase transition can occur. 
In order to prove this explicitly, we will use a mean-field approach to study the full Hamiltonian in \Eq{HF}. We shift the bosonic fields as  
\begin{eqnarray}
\label{AVG}
\mc{r}=\psi_{\mathbf{r}} + \fc{r},\quad
\langle \mc{r}\rangle=\psi_{\mathbf{r}},\quad
\langle \fc{r}\rangle=0,
\end{eqnarray}
where the expectation value is taken over the ground state.
In order to decouple the different sites, we disregard the coupling between fluctuations at different sites and obtain, on each site $\mathbf{r}$, the following Hamiltonian
\begin{eqnarray}
\label{HMF}
H^{\text{MF}}_{\mathbf{r}}&=& \bar{\omega}\fc{r}\fa{r}+  \frac{\omega_{12}}{2} \sigma^z_{\mathbf{r}} 
+ \Omega(\psi_{\mathbf{r}}  + \psi_{\mathbf{r}} ^*+\fc{r}+\fa{r}) \sigma^x_{\mathbf{r}}  \nonumber \\&&
+\sum_{\mathbf{r'}} \frac{ \tilde{\omega}_{\lvert \mathbf{r-r'}\lvert}}{\sqrt{N}} (\psi_{\mathbf{r}} \psi_{\mathbf{r'}}^*   + \psi_{\mathbf{r'}} \fc{r}+\psi_{\mathbf{r'}} ^*\fa{r}).
\end{eqnarray}

We expect the previous mean field Hamiltonian to present a delocalized superradiant phase transition  \cite{Tureci12} for large enough values of the inter-site coupling. 
In order to allow for the field to condense in any of the original $\mathbf{q}$ modes, we chose the following inhomogeneous ansatz for the mean field parameter
\begin{eqnarray}
\label{ansatz}
\begin{array}{rcll}
\psi_{\mathbf{r}}^{\mathbf{q}} &=& \psi \sqrt{2}\cos(\mathbf{qr})& \text{ for } \mathbf{q}\neq \mathbf{0} \\ 
\psi_{\mathbf{r}}^{\mathbf{q}} &=& \psi &\text{ for } \mathbf{q}=\mathbf{0},
\end{array}
\end{eqnarray}
with $\psi$ real and the $q$-dependent normalization is needed to have a continuous energy for the bosonic field (that depends on the integral of $\psi_{\mathbf{r}}^{\mathbf{q}}$ over $\mathbf{r}$).

Inserting the ansatz in \Eq{ansatz} into \Eq{HMF} we obtain
\begin{eqnarray}
\label{HMFL}
H^{\text{MF}}_{\mathbf{r}}&=& \omega_q  \psi_{\mathbf{r}}^{\mathbf{q}\,2} +\frac{\omega_{12}}{2} \sigma_{\mathbf{r}}^z+ 2\Omega \psi_{\mathbf{r}} \sigma_{\mathbf{r}}^x 
\\&& +  
\bar{\omega} \fc{r}\fa{r}+  (\Omega \sigma_{\mathbf{r}}^x+ \psi_{\mathbf{r}}^{\mathbf{q}}  \omega_q)( \fc{r}+\fa{r})\nonumber.
\end{eqnarray}

\begin{figure}[h!]
\begin{center}
\includegraphics[width=7.5cm]{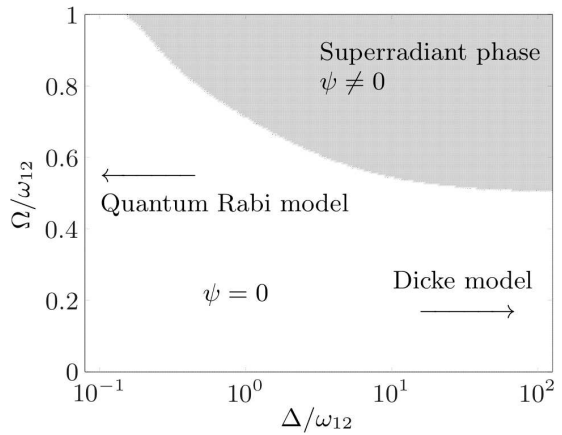}
\caption{\label{phaseboundary} Phase diagram of the considered flat band model at the resonance $\omega_m=\omega_{12}$, as a function of the normalized coupling,
$\frac{\Omega}{\omega_{12}}$, and of the normalized half width of the bosonic dispersion, $\frac{\Delta}{\omega_{12}}$. We see how the normalized coupling at the critical point
is $0.5$ in the limit of wide bands, as predicted  by the Dicke model and diverges for flat optical dispersions, as the system becomes well described by a quantum Rabi model.}
\end{center}
\end{figure}
In order to calculate the phase diagram of the system we have to find the values of $\psi$ and $q$ that minimize the ground state energy of the
total mean field Hamiltonian  $H^{\text{MF}}_{\text{FBM}}=\sum_{\mathbf{r}} H^{\text{MF}}_{\mathbf{r}}$. 
As a first step we notice that  the Hamiltonian in \Eq{HMFL}  is invariant over a contemporary sign exchange of $\psi_{\mathbf{r}}^{\mathbf{q}}$, $\sigma_{\mathbf{r}}^x$ and $ \fc{r}$. 
 The expectation value of an arbitrary observable $\eta_{\mathbf{r}}$, over site ${\mathbf{r}}$, can thus be written, close enough to the transition to be able to keep only the lowest order term in the order parameter, in the form
 \begin{eqnarray}
 \label{locob}
\eta_{\mathbf{r}}=A_q+B_q  \psi_{\mathbf{r}}^{\mathbf{q}\,2},
\end{eqnarray}
where $A_q$ and $B_q$ are independent of $ \psi_{\mathbf{r}}^{\mathbf{q}}$ (but they can still depend upon $q$ through $\omega_q$).
An homogeneous observable, $\eta$, is equal to the average, over all sites ${\mathbf{r}}$, of local observables $\eta_{\mathbf{r}}$. From Eqs. (\ref{ansatz}) and (\ref{locob}), we thus have that 
 \begin{eqnarray}
 \label{globob}
\eta=A_q+B_q \psi ^2. 
\end{eqnarray}
For the sake of determining the phase boundary, we can thus consider the homogeneous case, dropping the site index.
Moreover, if we calculate the ground state expectation value, we obtain from  \Eq{AVG} 
\begin{eqnarray}
\label{HMFLAVG}
\langle H^{\text{MF}} \rangle &=&\omega_q  \psi^2+ 
\langle \frac{\omega_{12}}{2} \sigma^z + 2\Omega \psi \sigma^x 
\nonumber \\&& +  
\bar{\omega} \fc{}\fa{}+  \Omega \sigma^x( \fc{}+\fa{})\rangle.
\end{eqnarray}
The first term in  \Eq{HMFLAVG} is a strictly increasing function of $\omega_q$, while the others do not depend upon $q$.
The system will thus always condense in the lowest lying bosonic mode and, now on, we can take $\omega_q=\omega_m$. 
We can thus limit ourselves to study a single, site independent local Hamiltonian
\begin{eqnarray}
\label{HMFLH}
H^{\text{MF}}&=&\omega_m  \psi^2 + \frac{\omega_{12}}{2} \sigma^z+ 2\Omega \psi \sigma^x \\&&+  
\bar{\omega} \fc{}\fa{}+  (\Omega \sigma^x+ \psi  \omega_m)( \fc{}+\fa{})\nonumber.
\end{eqnarray}

An analytic limit for the phase boundary can be obtained for broad dispersions (such that the average energy mismatch is much larger than the coupling, $\Delta \gg \Omega$). 
In this limit, we see from the fourth term in \Eq{HMFLH} that the energy cost of fluctuations increases, eventually completely freezing them. 
It is thus possible to neglect quantum fluctuations and consider only the coherent part of the field
\begin{eqnarray}
\label{HMA}
\lim_{\bar{\omega} \rightarrow \infty} H^{\text{MF}}&=&\omega_m  \psi^2 + \frac{\omega_{12}}{2} \sigma^z+ 2\Omega \psi \sigma^x.
\end{eqnarray}
The previous Hamiltonian can be analytically diagonalized, yielding an energy, close to the QPT boundary 
$E=(\omega_m-\frac{4\Omega^2}{\omega_{12}})\psi^2,$
leading to a second order QPT for the critical value 
\begin{eqnarray}
\label{cond} 
\Omega=\frac{\sqrt{\omega_{m} \omega_{12}}}{2},
\end{eqnarray}
that is the critical value we expect from the Dicke model \cite{Emary03,Lambert04}.
This result justifies, {\it a posteriori}, the choice to model flat band systems with steep enough bosonic dispersions as Dicke models. 
In this case, all the bosonic modes except the almost resonant one eventually become completely nonresonant and can thus be neglected.
 
For intermediate values of  $\Delta$, we diagonalized $H^{\text{MF}}$  in a truncated Fock space for different values of $\psi$, finding then the value that minimizes the ground state energy \cite{Greentree06,Koch09}.
In Fig. \ref{phaseboundary} we plot the phase diagram as a function of the normalized coupling, $\frac{\Omega}{\omega_{12}}$, and of the normalized half width of the bosonic dispersion, $\frac{\Delta}{\omega_{12}}$, for the resonant case $\omega_m=\omega_{12}$.
In the two limiting cases of $\frac{\Delta}{\omega_{12}}$ going toward zero and infinity, we find results consistent with our previous analytical estimations: the critical point diverges toward infinite values of the normalized coupling when the dispersion is almost flat (consistently with the fact that the quantum Rabi model does not present a critical point) or it converges to $0.5$, accordingly to \Eq{cond}, for very broad dispersions.

\begin{figure}[h!]
\begin{center}
\includegraphics[width=7.5cm]{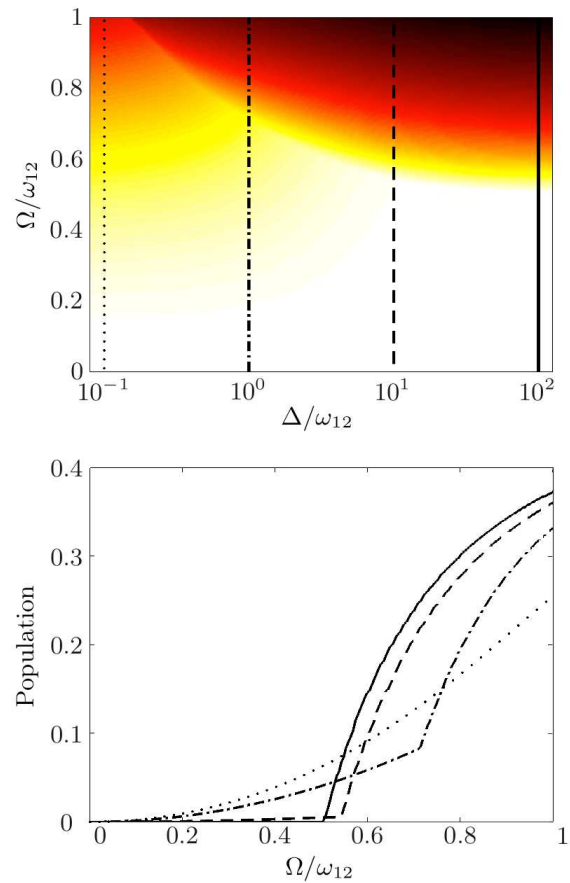}
\caption{\label{population}
Average occupation of the states in the excited band, for $\omega_m=\omega_{12}$. In the top panel we show a colorplot as a function
of the normalized coupling, $\frac{\Omega}{\omega_{12}}$, and of the normalized half width of the bosonic dispersion, $\frac{\Delta}{\omega_{12}}$.
In the lower panel we show instead four cuts of the same image, along the four vertical lines visible in the colorplot. We see how the system interpolates
between the continuous behavior of the quantum Rabi model and a second order quantum phase transition for the Dicke model.}
\end{center}
\end{figure}

In Fig. \ref{population} we show the average occupation of  the excited band, 
$\frac{1}{N}\sum_k \langle \ccm{2}{k}\cam{2}{k} \rangle=\langle \sigma_z \rangle+\frac{1}{2}$, 
for the same parameters. From the bottom panel of Fig. \ref{population} it is evident the continuous transition between the Dicke-like quantum phase transition  for broad dispersions (solid line)
and the Rabi-like crossover  (dotted line).

In conclusion, we investigated the physics of a model describing a multimode bosonic field  coupled to two flat electronic bands.
We proved that, if the bosonic mode is also dispersionless, the presented model maps exactly into an ultrastrongly coupled quantum Rabi model, allowing to simulate such a model with controllable coupling strengths. Using a mean field theory, we  showed that, engineering the dispersion of the bosonic field, the physics of such a model can be continuously tuned from a quantum Rabi model (with no critical point)  to a Dicke model exhibiting a superradiant QPT. % In particular, our results suggest that proposals to observe photonic superradiant QPT in flat band models \cite{Hagenmuller12} need to carefully take into account the full photonic environment in order to determine the possibility to reach the quantum critical point.

We thank  D. Hagenm\" uller, J. Keeling, P. Nataf and E. Solano for useful discussions.

\end{document}